# Tensile fracture behavior of short carbon nanotube reinforced polymer composites: a coarse-grained model

Behrouz Arash[1,*], Harold S. Park[2], Timon Rabczuk[1,3,†]

[1]*Institute of Structural Mechanics, Bauhaus Universität-Weimar, Marienstr 15, D-99423 Weimar, Germany*

[2]*Department of Mechanical Engineering, Boston University, Boston, Massachsetts 02215, USA*

[3]*School of Civil, Environmental and Architectural Engineering, Korea University, Seoul, Republic of Korea*

**Abstract**

Short-fiber-reinforced polymer composites are increasingly used in engineering applications and industrial products owing to their unique combination of superior mechanical properties, and relatively easy and low-cost manufacturing process. The mechanical behavior of short carbon nanotube (CNT) polymer composites, however, remains poorly understood due to size and time limitations of experiments and atomistic simulations. To address this issue, the tensile fracture behavior of short CNT reinforced poly (methyl methacrylate) (PMMA) matrix composites is investigated using a coarse-grained (CG) model. The reliability of the CG model is demonstrated by reproducing experimental results on the strain-stress behavior of the polymer material. The effect of the nanotube weight fraction on the mechanical properties, i.e. the Young's modulus, yield strength, tensile strength and critical strain, of the CNT/polymer composites is studied in detail. The dependence of the mechanical properties of the composites on the orientation and length-to-diameter aspect ratio of nanotube reinforcements is also examined.

---

[*] Author to whom correspondence should be addressed. E-mail address: behrouz.arash@uni-weimar.de, Tel: +49 3643 584511.

[†] Author to whom correspondence should be addressed. E-mail address: timon.rabczuk@uni-weimar.de, Tel: +49 3643 584511.

## 1. Introduction

Short-fiber-reinforced polymers (SFRPs) are a developing class of composite materials, which have attracted intense attention due to their outstanding mechanical, thermal and electrical properties [1, 2]. The mechanical properties of SFRP composites can reach stiffness levels attainable with continuous fibers, while the ability of unreinforced polymers to be formed into complex shapes is also preserved [3]. In addition, the production process for short fiber composites is more economical than continuous fiber composites [4]. This compromise between cost and performance makes short fiber polymer composites as excellent alternatives in electronic, automotive, oilfield and chemical industries [5]. Among different types of fibers used in the composites, the outstanding mechanical properties of carbon nanotubes (CNTs) promise ultra-high-strength reinforcements in high-performance polymer matrix composites [6].

In order to develop SFRPs, a quantitative understanding of their mechanical properties is of great significance since the mechanical performance of the materials strongly depends on controllable parameters such as fiber length and orientation [7]. Up to now, a number of experimental studies have been conducted in the literature on the mechanical behavior of reinforced polymer composites [8-12]. The experimental investigations, however, provide insufficient insight into molecular scale processes such as the interfacial interactions between nanotubes and matrix. The main reason for the drawback is rooted in the limited resolution of experimental techniques at the nanoscale. Furthermore, experimental efforts frequently encounter difficulties in fabricating SFRPs with uniformly distributed fibers with desired sizes.

Molecular dynamics (MD) simulations, in contrast, provide detailed information of phenomena at the nanoscale such as stick–slip mechanisms and the interfacial interactions between matrix and reinforcements [13-16]. MD simulations also facilitate the interpretation of experimental data, and additionally open a route to new designs of nanocomposites. Therefore, molecular simulations are necessary in the understanding of mechanical behavior of reinforced polymer nanocomposites. Zhu et al. [17] studied the elastic properties of an epoxy Epon 862 matrix with a size of 4.028×4.028×6.109 nm$^3$ reinforced by short (10, 10) CNTs with length-to-diameter aspect ratios of 2.15 and 4.5. They reported that short CNT fibers increase the Young's modulus of the polymer matrix up to 20% compared to the pure Epon 862 matrix. Molecular simulation studies on the elastic properties of short single-walled CNT (SWCNT) reinforced Poly (vinylidene fluoride) (PVDF) matrix composites [18] showed that a (5, 5) SWCNT with a length of 2 nm can increase the Young's modulus of a CNT/PVDF unit cell by 1 GPa. The simulation unit cell consists of a (5, 5) SWCNT with a volume fraction of 1.6% embedded in 60 PVDF chains. Arash et al. [19] investigated the mechanical behavior of CNT/poly (methyl methacrylate) (PMMA) polymer composites under tension. They proposed a method for evaluating the elastic properties of the interfacial region of CNT/polymer composites. Their simulation results on the elastic properties of a PMMA polymer matrix with a size of 3.7×3.7×8 nm$^3$ reinforced by a short (5, 5) SWCNT reveal that the Young's modulus of the composite increases from 3.9 to 6.85 GPa with an increase in the length-to-diameter aspect ratio of the nanotube from 7.23 to 22.05.

Although molecular simulations have been broadly used in modeling reinforced polymer nanocomposites, the massive computational effort required by the simulations strictly limits their applicability to small molecular systems over a limited time scale. These drawbacks hinder MD simulations to study the effect of fiber sizes and orientations on the mechanical behavior of reinforced polymer nanocomposites. In order to overcome these limitations, coarse-grained (CG) models beyond the capacity of molecular simulations have been developed in the literature [20-22]. The principle of CG models is to map a set of atoms to a CG bead, which reduces the atomistic degrees of freedom, resulting in substantial increases in the accessible time and length-scales while partially retaining the molecular details of an atomistic system. Up to now, many CG models have been developed for polymer materials [21, 23, 24]. Recently, the reliability of these approaches in modeling graphenes and CNTs has been also examined [25-28]. Ruiz et al. [27] established a CG model for the elastic and fracture behavior of graphenes with a ~200 fold increase in computational speed compared to atomistic simulations. Zhao et al. [28] calibrated parameters of the CG stretching, bending and torsion potentials for SWCNTs to study their static and dynamic behaviors. They also derived parameters of non-bonded van der Waals (vdW) interactions between CNTs in a bundle. The CG model was shown to have great potential in the analysis of the mechanical properties of CNT bundles with low computational cost compared to atomistic simulations. Arash et al. [29] developed a comprehensive CG model of CNT reinforced polymer composites capturing the non-bonded interactions between polymer chains and nanotubes. They then employed the model to study the elastic properties of short and long CNT reinforced PMMA polymer composites. Despite the simulation studies on the elastic properties of short CNT reinforced polymer composites, there is still no simulation investigation on the fracture behavior of a polymer matrix reinforced by randomly distributed short CNTs. Furthermore, the effects of nanotubes length and orientation on the mechanical properties of short CNT/polymer composites at large deformations have not been quantified. Hence, a quantitative study on the mechanical properties of short CNT/polymer composites is essential to achieve a successful design, synthesis, and characterization of the nanocomposites.

In this study, the mechanical behavior of short CNT/PMMA composites under tension is investigated in elastic and plastic regimes using a CG model. The applicability of the CG model in predicting the strain-stress behavior of PMMA polymer is examined using experimental results reported in the literature. The effects of the weight fraction, orientation and length-to-diameter aspect ratio of unidirectional and randomly distributed CNT reinforcements on the mechanical properties of the nanocomposites are studied in detail. The mechanical behavior of the CNT/PMMA composites observed in the CG simulations is also interpreted using a micromechanical continuum model.

## 2. Methods

In this study, we used a CG model that was previously developed and examined for modeling CNT/PMMA polymer composites [29]. In this model, each methyl methacrylate ($C_5O_2H_8$) monomer is treated as a bead with an atomic mass of 100.12 amu as illustrated in Fig. 1 (a). The center of the bead is chosen to be the center of mass of the monomer. The pseudoatom is defined as P bead, which enables a 15 fold decrease in the number of degrees of freedom (DOF) of a polymer chain compared to its corresponding full atomistic system. The CG model was also developed for modeling (5, 5) CNTs, where each five atomic rings are mapped into a CG bead with an atomic mass of 600.55 amu (see Fig. 1(b)). The bead is defined as a C bead, which decreases the number of DOF of a nanotube 50 fold with respect to full atomistic simulations.

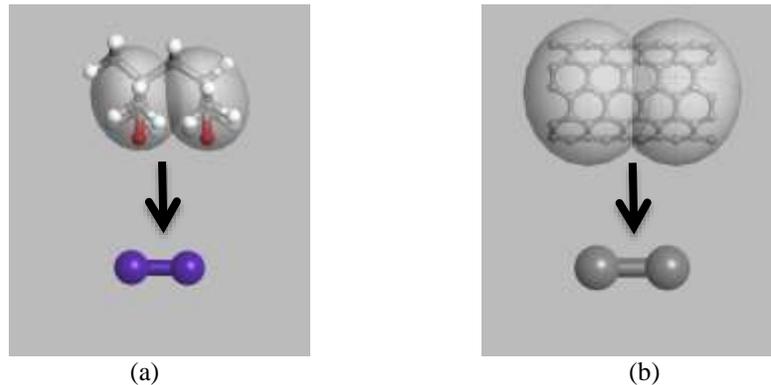

(a) (b)

Fig. 1. (a) Two monomers of a PMMA polymer chain and its CG model made of two P beads, and (b) a (5, 5) CNT with 10 rings of carbon atoms and its CG model made of two C beads.

The CG force field is decomposed into bonded and non-bonded potential functions. The total potential energy, $E_{total}$, of a system is therefore written as the sum of energy terms associated with the variation of the bond length, $E_b$, the bond angle, $E_a$, the dihedral angle, $E_d$, the vdW interactions, $E_{vdW}$, and the constant free energy of the system, $U_0$, as $U_{total} = \sum_i E_{b_i} + \sum_j E_{a_j} + \sum_k E_{d_k} + \sum_{lm} E_{vdW_{lm}} + U_0$. The functional forms of the contributing terms for a single interaction are as follows:

$$E_b(d) = \frac{k_d}{2}(d-d_0)^2 \quad \text{for} \quad d < d_{cut}, \tag{1a}$$

$$E_a(\theta) = \frac{k_\theta}{2}(\theta-\theta_0)^2, \tag{1b}$$

$$E_d(\phi) = \frac{k_\phi}{2}[1 + \cos 2\phi], \tag{1c}$$

$$E_{vdW}(r) = D_0\left[\left(\frac{r_0}{r}\right)^{12} - 2\left(\frac{r_0}{r}\right)^6\right], \tag{1d}$$

where $k_d$ and $d_0$ are the spring constant of the bond length and the equilibrium bond distance, respectively; $k_\theta$ and $\theta_0$ are respectively the spring constant of the bond angle and the equilibrium bond angle; $k_\phi$ and $\phi$ are the spring constant of the dihedral angle and the dihedral angle, respectively. $D_0$ and $r_0$ are the Lennard-Jones parameters associated with the equilibrium well depth and the equilibrium distance, respectively. A potential cutoff of 1.25 nm is used in calculation of vdW interactions. The parameters of the force field are listed in Table 1. The following simulations were performed using Accelrys Materials Studio 7.0.

Table 1. Parameters of the CG force field for C and P beads.

| Type of interaction | Parameters | C bead | P bead | C-P beads |
|---|---|---|---|---|
| **Bond** | $K_0(kcal/mol/Å^2)$ | 1610.24 | 194.61 | - |
|  | $d_0(Å)$ | 6.03 | 4.02 | - |
|  | $d_{cut}(Å)$ | 6.77 | 4.3 | - |
| **Angle** | $K_\theta(kcal/mol/Å^2)$ | 66148.01 | 794.89 | - |
|  | $\theta_0(°)$ | 180 | 89.6 | - |
| **Dihedral** | $K_\phi(kcal/mol)$ | 14858.80 | 42.05 | - |
| **vdW** | $D_0(kcal/mol)$ | 10.68 | 1.34 | 2.8 |
|  | $r_0(Å)$ | 9.45 | 6.53 | 7.71 |

## 3. Results and discussion

### 3.1. Mechanical properties of polymer matrix

To evaluate the applicability of the CG model in predicting the mechanical properties of CNT/PMMA composites, the strain-stress behavior of PMMA polymer is first investigated. Results obtained by the CG model are then justified by experimental data available in the literature. A simulation unit cell with a size of 12×12×12 nm$^3$ and periodic boundary conditions that contains amorphous PMMA polymer with a mass density of 1.1 $g/cm^3$ is initially constructed. Each polymer chain is composed of 100 repeated monomer units. The CG representative volume element (RVE) consists of 11400 beads, which is equivalent to a full atomistic system with 171000 atoms.

In order to find a global minimum energy configuration, a geometry optimization is first performed using the conjugate-gradient method [30]. The system is then allowed to equilibrate over the isothermal–isobaric ensemble (NPT) ensemble at room temperature of 298 K and atmospheric pressure of 101 kPa for 10 ns. In the NPT simulations, the time step is set to be 10 fs. The Andersen feedback thermostat [31] and the Berendsen barostat algorithm [32] are respectively used for the system temperature and pressure conversions. The NPT simulation is followed by a further energy minimization. The process removes internal stresses in the polymer system. After the preparation of the system, the constant-strain minimization method is applied to the equilibrated system to measure the material properties of the composite. A small tensile strain of 0.02% is applied to the periodic structure shown in Fig. 2 (a) in the *x*-direction. The application of the static strain is accomplished by uniformly expanding the dimensions of the simulation cell in the loading direction and re-scaling the new coordinates of the atoms to fit within the new dimensions. After each increment of the applied strain, the potential energy of the structure is re-minimized keeping the lattice parameters (i.e. simulation box sizes and angles) fixed. The stress in the unit cell is then calculated based on the virial stress definition. During the static deformation, the pressure in *y*- and *z*-directions is kept at atmospheric pressure by controlling the lateral dimensions. This process is repeated for a series of strains from which the variation of stress versus applied strain is obtained.

Fig. 2 (b) presents the true strain-true stress behavior of the polymer system subjected to the static tensile loading in the *x*-direction predicted by the CG model. The strain-stress response of the material is also compared to experimental results under quasi-static loading. The Young's modulus of the PMMA polymer predicted by the CG model is calculated to be 2.98 GPa, which is in fair agreement with the available experimental results varying from 2.24 to 3.27 GPa [33, 34]. From Table 2, the yield strength of the PMMA polymer obtained from the present CG model and experiments with strain rates of $2.42\times10^{-4}$ and $9.87\times10^{-5}$ [34] are respectively 52.30, 56.6 and 51.4 MPa, showing a percentage difference up to 7% with experimental data. The tensile strength of PMMA polymer predicted by the CG model is calculated to be 85.82 MPa, which is in good agreement with those measured in experiments [34], i.e. 86.9 MPa with the strain rate of $2.42\times10^{-4}$ and 79.6 MPa with the strain rate of $9.87\times10^{-5}$. As listed in Table 1, the critical strain of the polymer is also measured to be 0.159, 0.165 and 0.178 using the CG model and experimental tests with the strain rates of $2.42\times10^{-4}$ and $9.87\times10^{-5}$ [34], respectively. The critical strain is defined as the strain at which the stress drops to zero as shown in Fig. 2 (b) for the CG model. Figs. 2 (c) and (d) illustrate the initiation and propagation of tensile fracture failure in the polymer system at the critical strain of 0.159 and strain of 0.198. The simulation results reveal that the CG model is able to effectively estimate the mechanical behavior of PMMA polymer.

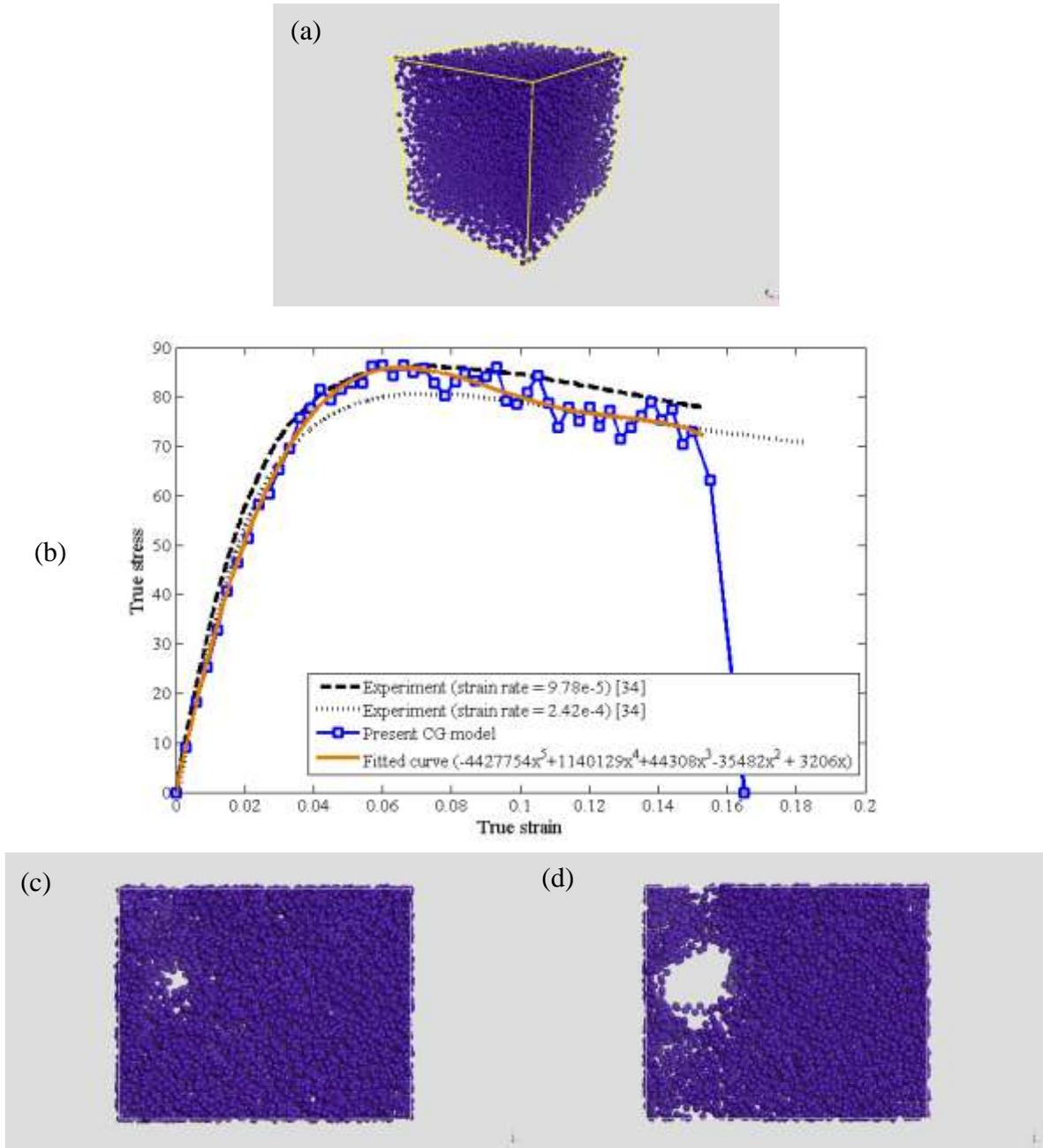

Fig. 2. (a) Initial configuration of an RVE of pure PMMA polymer with a size of 12×12×12 nm$^3$ (b) true stress-true strain behavior of the polymer system subjected to a uniaxial tensile loading, (c) snapshot at the critical strain of 0.159, and (d) snapshot at strain of 0.198. The RVE contains 11400 beads, which are equivalent to 171000 atoms.

Table 2. Mechanical properties of pure PMMA polymer matrix.

|  | Experiment (Strain rate=2.42e-4) | Experiment (Strain rate=9.87e-5) | Present CG model |
|---|---|---|---|
| **Young's modulus (GPa)** | 3.27 | 3.12 | 2.98 |
| **Yield strength (MPa)** | 56.6 | 51.4 | 52.30 |
| **Tensile strength (MPa)** | 86.9 | 79.6 | 85.82 |
| **Critical strain** | 0.165 | 0.178 | 0.159 |

### 3.2. Mechanical properties of unidirectional short nanotube polymer composites

Following the successful application of the CG model in predicting the mechanical behavior of PMMA polymer, we extend the CG model for the failure prediction of unidirectional short CNT reinforced PMMA composites. A unit cell of PMMA polymer with the same size of 12×12×12 nm$^3$ that contains four unidirectional (5, 5) CNTs is initially constructed as illustrated in Fig. 3 (a). The length of CNTs is 10 nm, and their weight fraction is set to be 0.03 (3 wt%). The unit cell contains 11164 beads, which is equivalent to a full atomistic system with 169700 atoms. In the following simulations, the mass density of CNT/PMMA polymer is set to be 1.1 $g/cc$. The CG RVE consists of 11164 beads, which is equivalent to a full atomistic system with 169700 atoms. The true strain-true stress curve of the CNT(3 wt%)/PMMA polymer composite is presented in Fig. 3 (b) from which the Young's modulus and tensile strength are respectively obtained to be 3.34 GPa and 95.50 MPa, showing percentage increases of 16 and 11% compared to pure polymer. From Fig. 3 (b), the tensile fracture failure in the CNT/PMMA composite occurs at the critical strain of 0.142 as illustrated in Fig. 3 (c). The fracture further propagates through the material at the strain of 0.19 as shown in Fig. 3 (d). In comparison to the pure polymer, the critical strain decreases as much as 10% in the presence of CNTs. The observation can be interpreted as the debonded CNT reinforcements behaving like voids at large strains, which weaken the composite.

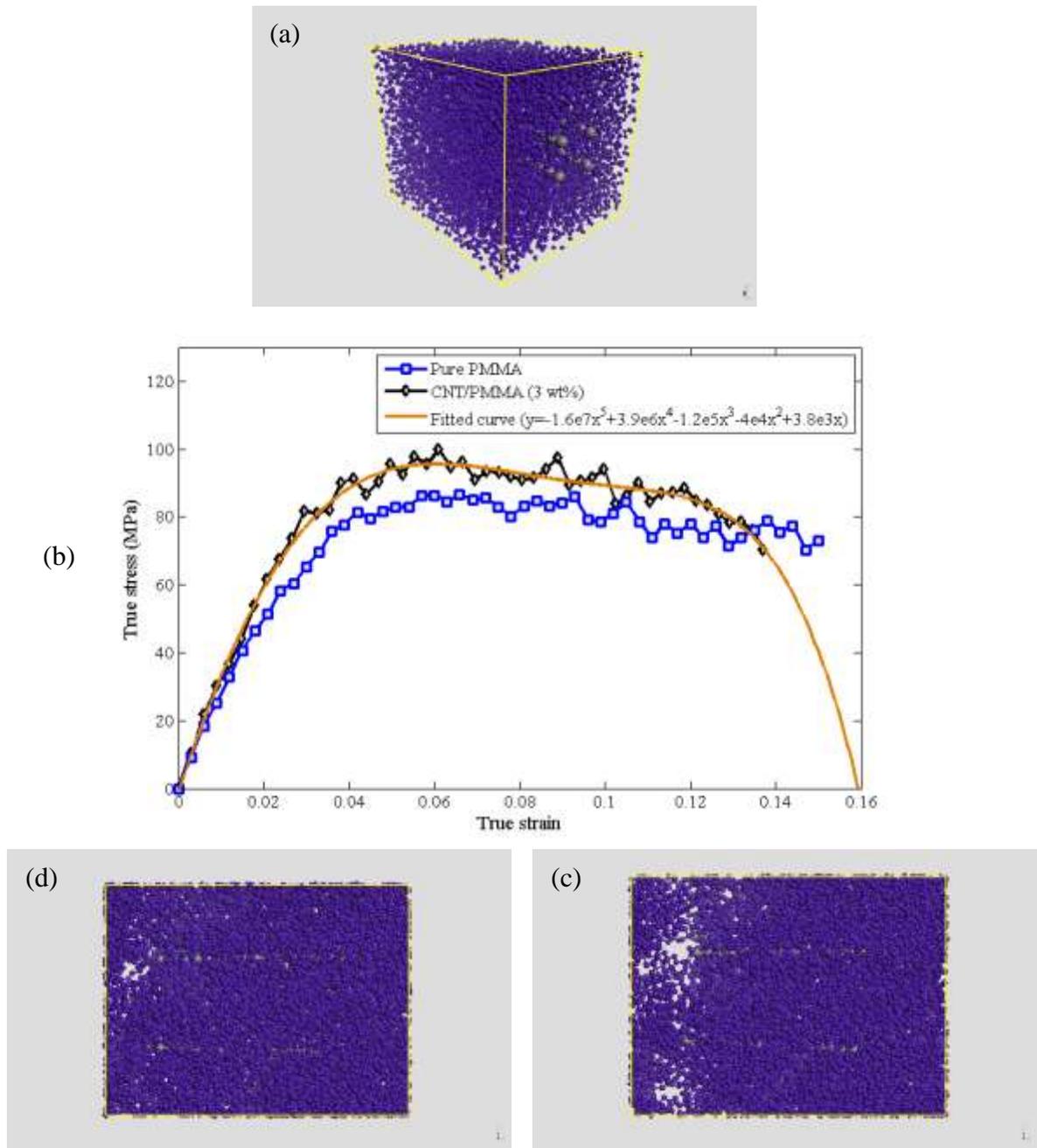

Fig. 3. Tensile fracture of a CNT (3 wt%)/PMMA composite with 10-nm long (5, 5) CNTs aligned in the load direction: (a) initial configuration of the composite (b) the true stress-true strain behavior, (c) snapshot at the critical strain of 0.142, and (d) snapshot at the strain of 0.19. The size of unit cell is $12 \times 12 \times 12$ nm$^3$ that contains 11164 beads. The number of beads is equivalent to 169700 atoms.

To further investigate the influence of short nanotube reinforcements on the material characteristics of the polymer composites, the effect of the weight fraction of CNTs on the mechanical behavior of CNT/PMMA composites is presented in Table 3. In the simulations, the RVE size, the size of nanotubes and the mass density of CNT/PMMA composite are the same as described in Fig. 3 (a). CNT fibers are unidirectional and their weight fraction varies from 3 to 10 wt%. To adjust the value of weight fraction, the number of (5, 5) CNTs in the polymer matrix differs from 4 to 12. From Table 3, the Young's modulus of CNT/PMMA composite increases from 3.34 to 3.63 GPa with an increase in the weight fraction of CNTs from 3 to 5 wt%, showing percentage increases of 16 and 26% compared to pure polymer, respectively. The Young's modulus is further raised to 4.22 and 4.52 GPa with increasing the weight fraction of CNTs to 8 and 10 wt%. It implies that short CNT fibers with the length-to-diameter ($L/D$) aspect ratio of 14.7 and weight fractions of 8 and 10 wt% significantly enhance the stiffness of a PMMA polymer matrix as high as 47 and 57%, respectively. The simulation results also show that the tensile strength of the CNT/PMMA composites increases from 95.50 to 100.72 MPa with increasing CNT weight fraction from 3 to 5 wt%, indicating percentage increases of 11 and 17% compared to pure polymer, respectively. The tensile strength of the CNT/PMMA composite with CNT weight fractions of 8 and 10 wt% respectively increases to 108.55 and 114.81 MPa. It implies that short CNTs with a weight fraction of 10 wt% aligned in the load direction noticeably strengthen the polymer matrix as much as 34%.

Table 3. Mechanical properties of CNT/PMMA polymer composites reinforced by 10-nm long (5, 5) CNTs aligned in the load direction.

| Weight fraction percentage (wt %) | Present CG model | | Krenchel's rule of mixtures | |
|---|---|---|---|---|
| | Young's modulus (GPa) | Tensile strength (MPa) | Young's modulus (GPa) | Tensile strength (MPa) |
| **Pure polymer** | 2.88 | 85.83 | - | - |
| 3 | 3.34 | 95.50 | 3.38 | 94.08 |
| 5 | 3.63 | 100.72 | 3.70 | 98.17 |
| 8 | 4.22 | 108.55 | 4.44 | 106.35 |
| 10 | 4.52 | 114.81 | 4.85 | 110.44 |

As presented in Fig. 4, the critical strain of CNT/PMMA composites increases from 0.148 at the weight fraction of 5 wt% to 0.170 at the weight fractions of 8 wt%, respectively. The critical strain further increases to 0.183 at the weight fraction of 10 wt%, indicating a percentage increase of 15% in comparison to the pure PMMA polymer. The simulation results reveal that CNT reinforcements with the weight fractions less than 5 wt% behave as voids at large strains, which decrease the critical strain of composites compared to the pure polymer material leading to a more brittle behavior. In contrast, CNT fibers with the weight fractions greater than 8 wt% improve the tear strength of polymer composites subjected to tensile loading by controlling the growth of cracks in the polymer matrix, leading to a more ductile behavior compare to pure polymer.

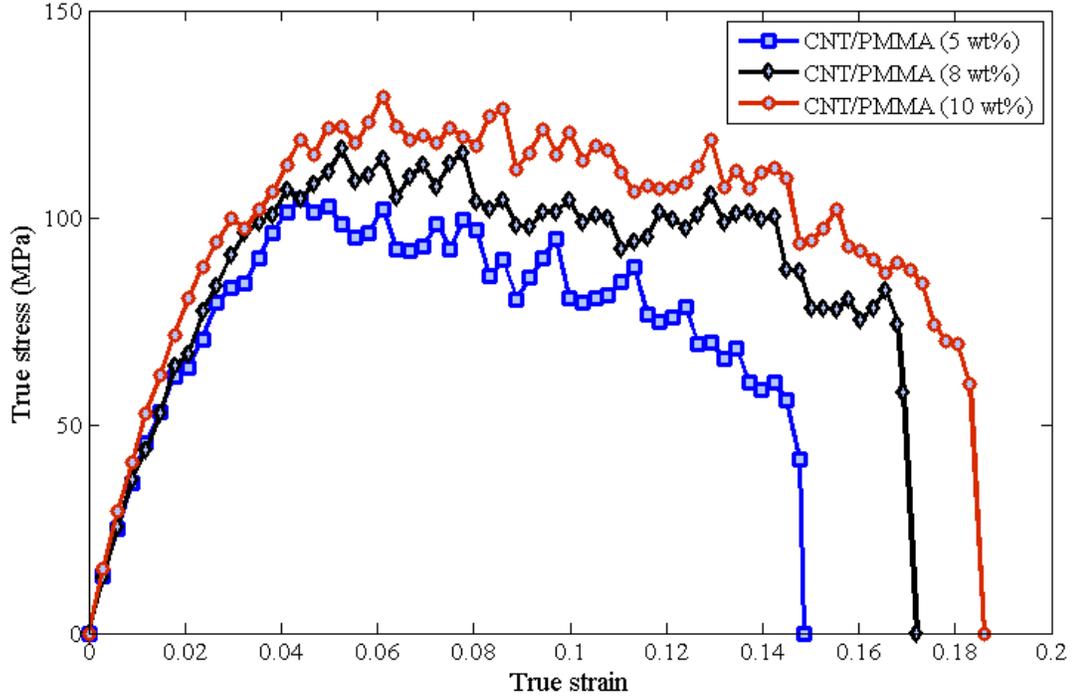

Fig. 4. Effect of the CNT weight fraction on the tensile fracture behavior of CNT/PMMA composites. The size of unit cell is 12×12×12 nm³ and the 10-nm long (5, 5) CNTs are aligned in the load direction.

The mechanical behaviors observed in the simulations can be interpreted by micromechanical continuum models. Micromechanics can be used to model the mechanical properties of composites. In this study, Krenchel's rule of mixtures for short fiber composites is used for estimating the Young's modulus of the polymer composites as [35]

$$E^c = \eta_0 \eta_l E^f V^f + E^m V^m, \tag{1}$$

where superindices $c, f$ and $m$ refer to composite, fiber and matrix, respectively. $E$ and $V$ denote the Young's modulus and volume fraction of the material. The Young's modulus of a (5, 5) CNT reinforcement is measured to be 1.65 TPa using MD simulations [19]. It is noteworthy that in calculation of the Young's modulus of the CNT, the nanotube is supposed to be a solid bar with a cross sectional area of $A_{CNT} = \frac{1}{4}\pi d^2$. The terms $\eta_l$ and $\eta_0$ represent the efficiency factors of fiber length and orientation. $\eta_0$ can be chosen to be equal to 1 for unidirectional fibers. The factor $\eta_l$ is given by [36]

$$\eta_l = 1 - \frac{\tanh\frac{\zeta L^f}{2}}{\frac{\zeta L^f}{2}}, \qquad (2)$$

where

$$\zeta = \frac{1}{r}\frac{E^m}{E^f(1-v)\ln\left(\frac{\pi}{4V^f}\right)^{1/2}} \qquad (3)$$

Here, $L^f$ and $r$ are the length and radius of fibers, and $v$ is the Poisson's ratio of matrix. In addition, the composite tensile strength, $\sigma^c$, is obtained as [37]

$$\sigma^c = \eta_0\eta_l\sigma^f V^f + \sigma^m V^m, \qquad (4)$$

in which $\sigma^f$ and $\sigma^m$ are the tensile strength of fiber and matrix, respectively. The Kelly-Tyson model [38] suggests that the maximum stress in fibers embedded in a matrix is proportional to the fiber aspect ratio ($L/D$) and the fiber/matrix interfacial shear stress ($\tau$). Incorporating this model into the rule of mixture gives the composite tensile strength as

$$\sigma^c = \eta_0\eta_l 2\tau(L/D)V^f + \sigma^m V^m. \qquad (5)$$

The interfacial shear stress between PMMA matrix and CNT has been obtained to be 35.9 MPa through pullout simulations [39].

Based on the above description, the Young's modulus and tensile strength of the CNT/PMMA composites calculated from Eqs. (1) and (5) are compared to those obtained from the CG model in Table 2. From Table 3, the Young's modulus of a PMMA matrix composite reinforced by 10-nm long (5, 5) CNTs with weight fractions of 3 and 5 wt% are respectively calculated to be 3.38 and 3.70 GPa using Eq. (1). The results reveal percentage differences of 1.1 and 1.9% at the weight fractions of 3 and 5 wt%. The percentage difference increases to 7.3% at the weight fraction of 10 wt%, where the Young's modulus of the CNT/PMMA composite predicted by the CG model and the micromechanical mode are 4.44 and 4.85 GPa, respectively. Furthermore, the tensile strength of the polymer composite calculated from Eq. (5) is 94.08 and 98.17 MPa at the weight fraction of 3 and 5 wt%, showing percentage differences of 1.5 and 2.5% compared to the CG simulation results. From Eq. (5), the tensile strength of the polymer composite is respectively obtained to be 106.35 and 110.44 MPa at the weight fraction of 8 and 10 wt%, revealing a percentage difference of up to 3.8% in comparison to the CG simulation results.

It is concluded that there is a good agreement between results obtained from the rule of mixture and CG simulations for unidirectional short CNT fibers. However, the micromechanical model is unable to predict the strain-stress behavior and the critical strain of the composite materials. Furthermore, the development of an accurate micromechanical model for predicting the mechanical properties of randomly distributed short fiber composites is quite difficult because of the complicated interactions at the interface of the fibers and matrix as well as the complex fiber length and orientation effects. Therefore, the value of the CG model lies in modeling polymer matrix composites reinforced by unidirectional and randomly oriented short CNTs, while accounting for the key interactions between polymer chains and the CNTs.

### 3.3. Mechanical properties of randomly distributed short nanotube polymer composites

The effective properties of a randomly distributed fiber composite material are obtained from a sufficiently large sample volume that is a statistical representative of the whole microstructure. Hill defined an RVE as a sample that (a) is structurally entirely typical of the whole mixture on average, and (b) contains a sufficient number of inclusions for the apparent overall moduli to be effectively independent of the surface values of traction and displacement [40]. In order to find an appropriate size for the RVE, we first fulfill a sample enlargement test. In simulations, cubic polymer matrices with side lengths ranging from 20 to 50 nm reinforced by randomly distributed (5, 5) CNTs are considered as illustrated in Fig. 5. The length of CNTs is 10 nm and their weight fraction is set to be 10 wt%. The mass density of the CNT/PMMA composite is also set to be 1.1 g/cc. In order to obtain quasi-isotropic mechanical properties, a uniform probability distribution function is used to position the CNTs inside the RVE. We compute the Young's modulus and tensile strength of an RVE with a side length of $l$. The calculations are continued for a larger RVE with a side length of $l' > l$. $l'$ is taken to be sufficiently large size of the RVE, if the following criteria are met [41]:

$$\frac{|E_{l'} - E_l|}{E'_l} < 0.01,$$

$$\frac{|\sigma_{cl'} - \sigma_{cl}|}{\sigma_{cl'}} < 0.01. \tag{6}$$

where $E_{l'}$ and $E_l$ are the Young's modulus of the RVE size of $l$ and $l'$, respectively, and $\sigma_{cl'}$ and $\sigma_{cl}$ are respectively the tensile strength of the RVE size of $l$ and $l'$.

From simulation results presented in Table 4, the Young's modulus of the CNT/PMMA composite is 3.35 and 3.27 GPa at the side lengths of 20 and 30 nm, respectively, showing a percentage difference of 2.4%. The tensile strength of the composite is also calculated to be 104.51 and 102.15 MPa at the side lengths of 20 and 30 nm, respectively, indicating a percentage difference of 2.3%. The percentage differences respectively decreases to 0.6 and 0.1% for the Young's modulus and tensile strength at the RVE size of 50×50×50 nm$^3$, which both meet the criteria defined in Eq. (6). At the RVE size, the Young's modulus and tensile strength of the CNT (10 wt%)/PMMA composite are obtained to be 3.24 GPa and 101.85 MPa, respectively. The number of beads in the RVE size is greater than 700,000, which is equivalent to a full atomistic system with more than 11 million atoms.

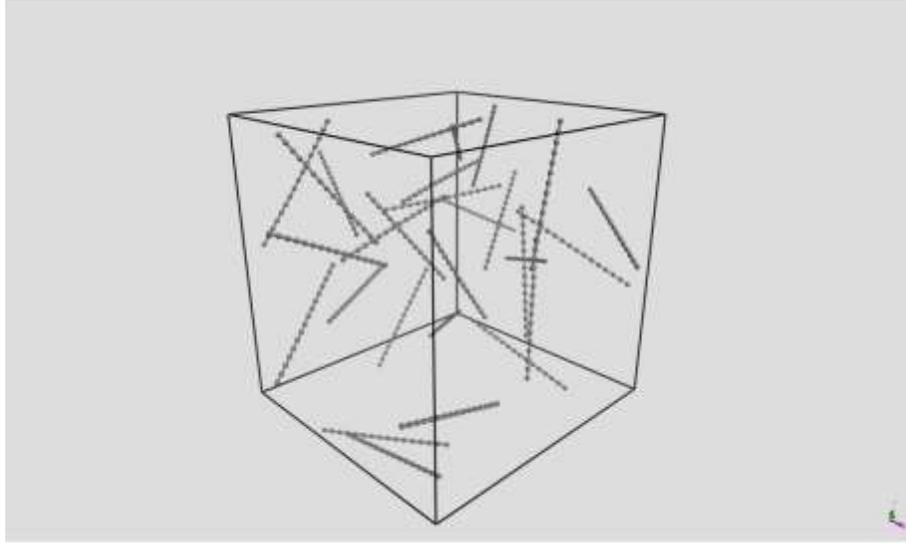

Fig. 5. A CNT (10 wt%)/PMMA composite with 10-nm long (5, 5) CNTs randomly distributed in three dimensional space. The size of unit cell is $20 \times 20 \times 20\ nm^3$ that contains 50648 beads. The number of beads is equivalent to 775400 atoms.

Table 4. Effect of RVE size on the mechanical properties of PMMA matrix reinforced by 10-nm long (5, 5) CNTs randomly distributed in plane. The weight ratio of CNTs is set to be 10 wt%.

| RVE size ($nm^3$) | Number of beads (number of equivalent atoms) | Young's modulus (GPa) | Tensile strength (MPa) |
|---|---|---|---|
| $20 \times 20 \times 20$ | 48520 (734,780) | 3.35 | 104.51 |
| $30 \times 30 \times 30$ | 163750 (2,479,882) | 3.27 | 102.15 |
| $40 \times 40 \times 40$ | 388160 (5,878,240) | 3.22 | 101.71 |
| $50 \times 50 \times 50$ | 756912 (11,462,568) | 3.24 | 101.85 |

After determining the appropriate RVE size, the effect of CNT weight fraction on the mechanical properties of randomly distributed CNT/PMMA composites is investigated. In the simulations, the size of CNTs and the mass density of the composite are the same as described in Fig. 5. As presented in Table 5, the Young's modulus of CNT/PMMA composite increases from 3.05 to 3.11 GPa with an increase in the weight fraction of CNTs from 5 to 8 wt%, respectively, revealing percentage increases of 6 and 8% in comparison to pure PMMA polymer. The Young's modulus further increases to 3.24 GPa at the CNT weight fraction of 10 wt%. It implies that randomly distributed short CNT fibers with the aspect ratio of $L/D = 14.7$ and the weight fraction 10 wt% improve the stiffness of a PMMA polymer matrix as much as 12.5%. From the simulation results, the tensile strength of the composite increases from 94.22 to 96.80 MPa with increasing the CNT weight fraction from 5 to 8 wt%, showing percentage increases of 7.7 and 10% compared to pure polymer. The tensile strength of the CNT/PMMA composite increases to 99.63 MPa at the weight fraction of 10 wt%, which implies that randomly distributed short CNTs with a weight fraction of 10 wt% strengthen the polymer matrix as high as 16%. It can be seen that while unidirectional 10-nm long (5, 5) CNTs significantly improve the mechanical properties of PMMA polymer composites (up to 57% for the Young's modulus and 34% for the tensile strength at the weight fraction of 10 wt%), the properties experience slight increases in the presence of the randomly distributed nanotubes (up to 12.5% for the Young's modulus and 16% for the tensile strength at the weight fraction of 10 wt%).

Table 5. Effect of CNT weight fraction on the mechanical properties of PMMA polymer matrix reinforced by randomly distributed (5, 5) 10-nm long CNTs. The RVE size is 50×50×50 nm³.

| wt (%) | Young's modulus (GPa) | Tensile strength (MPa) |
|---|---|---|
| **5** | 3.05 | 92.44 |
| **8** | 3.11 | 94.80 |
| **10** | 3.24 | 99.63 |

To further study the mechanical behavior of CNT/PMMA composites, the effect of the aspect ratio of CNTs ($L/D$) on their mechanical properties is presented in Table 6. The size of RVE is assumed to be 50×50×50 nm3 with and periodic boundary conditions are imposed in all directions. The diameter of the (5, 5) CNTs is 0.68 nm and their aspect ratio varies from 14.7 to 44.1, while their weight fraction is kept to be 8 wt%. The mass density of PMMA polymer matrix is set to be 1.1 g/cc. From Table 5, the Young's modulus of PMMA polymer matrix reinforced by (5, 5) CNTs increases from 3.11 GPa to 3.59 and 4.01 GPa with an increase in the aspect ratio of the nanotubes from 14.7 to 29.4 and 44.1. The simulation results demonstrate percentage increases of 15 and 29% in the stiffness of the CNT/PMMA composite with an increase in the aspect ratio of nanotubes from 14.7 to 29.4 and 44.1. The Young's moduli obtained for CNT/PMMA composites with CNTs aspect ratios of 29.4 and 44.1 are respectively 25 and 39% stiffer than a pure PMMA polymer material. Furthermore, the tensile strength of the CNT/PMMA composite is raised from 94.80 MPa to 101.82 and 111.61 MPa with increasing the aspect ratio of CNTs from 14.7 to 29.4 and 44.1. The results reveal that short CNT fibers with aspect ratios of 29.4 and 44.1 enhance the strength of PMMA polymer material as much as 19 and 30%, respectively. The higher composite stiffness and strength observed in simulations is rooted in the strengthening of CNT/polymer interfacial bonding with an increase in the aspect ratio of CNT fibers, which results in the increase of stress transfer between nanotubes and polymer chains.

Table 6. Effect of CNTs length-to-diameter aspect ratio on the mechanical properties of PMMA polymer matrix reinforced by randomly distributed nanotubes with a weight fraction of 8 wt%. The RVE size is 50×50×50 nm$^3$.

| $L/D$ | Young's modulus (GPa) | Tensile strength (MPa) |
|---|---|---|
| **14.7** | 3.11 | 94.80 |
| **29.4** | 3.59 | 101.82 |
| **44.1** | 4.01 | 111.61 |

## 4. Conclusions

The tensile fracture behavior of PMMA polymer matrix reinforced by short (5, 5) SWCNTs is investigated using a CG model. The strain-stress behavior of the polymer material predicted by the CG model is verified with experimental results available in the literature, revealing an excellent agreement between results obtained from the CG model and experiments. The effect of the CNT weight fraction on the Young's modulus, yield strength, tensile strength and critical strain of the polymer composites is studied. The simulation results show that unidirectional CNTs can significantly enhance the mechanical properties of the nanocomposites. The CG results demonstrate that the Young's modulus and tensile strength of a PMMA matrix reinforced by 10-nm long (5, 5) CNTs respectively increase as much as 57% and 34% compared to the pure polymer at the CNT weight fraction of 10 wt%. CNT reinforcements with the weight fractions greater than 8 wt% increase the critical strain of CNT/PMMA composites compared to the pure polymer material, leading to a more ductile behavior. The mechanical properties of PMMA polymer matrix reinforced by randomly distributed CNTs are studied and the appropriate RVE size, representing the whole microstructure, is obtained. The effects of the weight fraction and the length-to-diameter aspect ratio of short CNTs randomly distributed in the polymer matrix on the mechanical properties of the CNT/PMMA composites are explored. The simulation results show that (5, 5) CNT reinforcements with aspect ratios of 29.4 and 44.1 and the same weight fraction of 8 wt% enhance the Young's modulus of PMMA composites by 25 and 39%, respectively. The strength of PMMA polymer matrix respectively increases as much as 19 and 30% by adding CNT fibers with aspect ratios of 29.4 and 44.1 and the same weight fraction of 8 wt%.

**Acknowledgments**

The authors thank the support of the European Research Council-Consolidator Grant (ERC-CoG) under grant "Computational Modeling and Design of Lithium-ion Batteries (COMBAT)".

**Figure captions**

Fig. 1. (a) Two monomers of a PMMA polymer chain and its CG model made of two P beads, and (b) a (5, 5) CNT with 10 rings of carbon atoms and its CG model made of two C beads.

Fig. 2. (a) Initial configuration of an RVE of pure PMMA polymer with a size of 12×12×12 nm$^3$ (b) true stress-true strain behavior of the polymer system subjected to a uniaxial tensile loading, (c) snapshot at the critical strain of 0.159, and (d) snapshot at strain of 0.198. The RVE contains 11400 beads, which are equivalent to 171000 atoms.

Fig. 3. Tensile fracture of a CNT (3 wt%)/PMMA composite with 10-nm long (5, 5) CNTs aligned in the load direction: (a) initial configuration of the composite (b) the true stress-true strain behavior, (c) snapshot at the critical strain of 0.142, and (d) snapshot at the strain of 0.19. The size of unit cell is 12×12×12 nm$^3$ that contains 11164 beads. The number of beads is equivalent to 169700 atoms.

Fig. 4. Effect of the CNT weight fraction on the tensile fracture behavior of CNT/PMMA composites. The size of unit cell is 12×12×12 nm$^3$ and the 10-nm long (5, 5) CNTs are aligned in the load direction.

Fig. 5. A CNT (10 wt%)/PMMA composite with 10-nm long (5, 5) CNTs randomly distributed in three dimensional space. The size of unit cell is $20 \times 20 \times 20\ nm^3$ that contains 50648 beads. The number of beads is equivalent to 775400 atoms.

**Table captions**

Table 1. Parameters of the CG force field for C and P beads.

Table 2. Mechanical properties of pure PMMA polymer matrix.

Table 3. Mechanical properties of CNT/PMMA polymer composites reinforced by 10-nm long (5, 5) CNTs aligned in the load direction.

Table 4. Effect of RVE size on the mechanical properties of PMMA matrix reinforced by 10-nm long (5, 5) CNTs randomly distributed in plane. The weight ratio of CNTs is set to be 10 wt%.

Table 5. Effect of CNT weight fraction on the mechanical properties of PMMA polymer matrix reinforced by randomly distributed (5, 5) 10-nm long CNTs. The RVE size is $50 \times 50 \times 50$ nm$^3$.

Table 6. Effect of CNTs length-to-diameter aspect ratio on the mechanical properties of PMMA polymer matrix reinforced by randomly distributed nanotubes with a weight fraction of 8 wt%. The RVE size is $50 \times 50 \times 50$ nm$^3$.